\documentstyle[12pt,moriond,epsfig]{article}
\begin{document}
\newcommand{\etal}{{\em et al.}}
\heading{SEARCHES FOR LOW-METALLICITY GALAXIES and PRELIMINARY RESULTS 
FROM THE KPNO INTERNATIONAL SPECTROSCOPIC SURVEY}

\author{John J. Salzer}
{Wesleyan University, Middletown, CT USA}

\begin{moriondabstract}
Searches for low-metallicity galaxies are reviewed, focusing mainly on
efforts to discover systems that are relevant for use in measuring the
primordial helium abundance.  Wide-field objective-prism surveys for blue 
or emission-line
galaxies have played a major role in this field.  Previous surveys and
search techniques are highlighted.  Preliminary results from our new 
survey for emission-line galaxies, the KPNO International Spectroscopic 
Survey (KISS), are presented.  KISS is substantially deeper than previous 
photographic surveys, and represents the next generation of wide-field
slitless spectroscopic surveys.
\end{moriondabstract}

\section{Introduction}

Wide-field surveys for blue and/or emission-line galaxies have contributed 
a great deal to our knowledge of the properties of active and star-forming
galaxies.  It is difficult to imagine where the study of galaxian activity
would be today if astronomers did not have access to surveys with such familiar 
names as Markarian, Tololo, Michigan, and Case.  These surveys
have played a significant role in our knowledge of dwarf galaxies in
general, and in the measurement of elemental abundances in dwarf galaxies
in particular.  I have been asked to review the history of objective-prism
searches for activity in galaxies, with a focus on their use in finding
low-metallicity galaxies, as well as to describe the status of a modern
version of the classical objective-prism survey --- KISS.  

\section{Searches for Low-Metallicity Galaxies}

\subsection{Motivation}

The detection and study of galaxian systems with low metallicity has
been an active area of research for nearly thirty years \cite{SS}.  Since
low elemental abundances are generally believed to correlate with a 
restricted star-formation history, the discovery of low-Z galaxies is
of interest because it allows for the study of potentially young (in a 
cosmological sense), or at least unevolved, systems.  Much is still
unknown about the process of galaxy formation, although current evidence
suggests that most large galaxies (or their constituent parts) formed before 
z = 1.  Hence the discovery of late-forming systems could introduce a serious 
wrinkle in the currently-favored formation scenarios.  A few convincing candidates 
for genuinely young systems are known (\cite{S91},\cite{TXT},\cite{LvZ}),  
although they appear to be rare.  The abundances in these young objects 
are quite low (1/20th -- 1/50th solar), and all are dwarf galaxies.  
The study of these galaxies, as well as older, more evolved dwarf galaxies, 
has the potential to teach us a great deal about how chemical evolution 
proceeds in low-mass systems.  Unfortunately, the number of extremely 
metal-deficient galaxies known is very small, despite several focused efforts to 
find more \cite{KS}.  The discovery of additional metal-poor galaxies would be of
tremendous value.

An important aspect of the search for low-metallicity galaxies is that
these relatively pristine systems provide some of the best available 
sources for measuring the primordial helium abundance, one of the few
observable quantities that allows us to constrain models of the Big Bang.
Despite the fact that the lowest-Z galaxies discovered to date clearly have
had {\it some} amount of stellar processing, they are the least chemically-evolved 
systems for which a direct, accurate estimate of the helium
abundance is available.  Hence, they have been the subject of much work in the
past (and present), and have given the search for low-metallicity galaxies
an increased level of importance.  A more complete discussion of the 
derivation of the primordial helium abundance estimate is given by
Izotov \cite{YI} and references therein.

\subsection{Search Strategies}

How would one go about trying to detect low-metallicity galaxies?  In
particular, what would one look for if trying to detect galaxies with
abundances less than the current record-holder for low-metallicity, I Zw 18 
\cite{SK}?  This is a valid question, since obtaining an accurate abundance 
measurement for a galaxy can require a substantial amount of large telescope 
time.  Pre-selecting targets which have a good chance of having lower adundances 
is thus advantageous.  Three items that aid in the search for low-metallicity 
galaxies are listed below.

{\bf Low luminosity.}  It has been known for some time that lower luminosity
galaxies tend to have lower metal abundances (e.g., \cite{SK2}).  Although
the so-called luminosity--metallicity relationship has substantial scatter,
and some known low-Z galaxies are not particularly low luminosity (\cite{TXT}),
starting with a sample of dwarf galaxies is a sensible first step.  

{\bf Emission-line galaxies.} Since the desire is to be able to measure accurate
abundances for both metal atoms (e.g., oxygen, nitrogen) and helium, it will
help tremendously if the galaxies exhibit nebular spectra (from at least a portion
of the galaxy).  Thus, pre-selecting targets which show strong emission lines
is an advantage.  Objective-prism surveys which select galaxies based on line 
emission (see below) are a natural place to find such targets.  An added advantage 
of using emission-line galaxy (ELG) survey lists is that such surveys tend to
preferentially detect dwarfs (\cite{S89}, see also Figure \ref{fig:abs}).

{\bf H$\alpha$-selected surveys.}  If one focuses attention on the 
contents of objective-prism surveys, as suggested, then additional thought needs
to be given to the detection method of the surveys and any biases inherent in
them.  Most previous line-selected surveys have used the [O III]$\lambda$5007
line for detecting ELGs.  However, this introduces a metallicity bias into
the selection process.  Truly low-Z galaxies (less than 1/50th solar) will have
weak [O III] lines.  It would be better to select candidates via their H$\alpha$
emission, since for low-Z galaxies the H$\alpha$ line will be the strongest
in the optical portion of the spectrum. 

\subsection{Previous Surveys}

Our review of searches for low-Z systems will be brief,
since there is not sufficient room to mention all previous surveys here.  
A more complete discussion can be found in Salzer et al. \cite{S98}.  For the 
reasons outlined above, objective-prism surveys will be emphasized.  
Other searches for low-Z objects have looked at nearby dwarf galaxies (e.g., 
\cite{SK2}), while many of the known low-metallicity galaxies have simply 
been discovered to be so by chance observations (e.g., \cite{S91}).  
Only the systematic surveys will be discussed here.

There are two methods used in objective-prism surveys to find galaxies with a high 
level of activity: (1) an excess of flux in the UV portion of the spectrum over
``normal" objects, and (2) the detection of one or more emission lines in the
spectrum of an object.  The famous Markarian survey \cite{MRK} used the UV-excess
method to find starburst and Seyfert galaxies.  Since this was the earliest
systematic survey, and since it covered such a large fraction of the sky (36\%),
objects cataloged by Markarian have contributed a great deal to the understanding 
of galaxian activity.  Although objects as faint as B = 17 were discovered by
Markarian, the completeness limit of the survey is only B = 15.2 \cite{MAZ}.

Subsequent to the Markarian survey, a number of groups carried out deeper 
objective-prism surveys which relied either partially or exclusively on the
detection of emission lines.  Most worked at blue wavelengths, and hence were sensitive
to [O III]$\lambda$5007.  Examples of purely line-selected surveys include the 
Tololo \cite{TOL} and Michigan (UM)\cite{UM} surveys, while others such as the 
Second Byurakan \cite{SBS} and Case \cite{CASE} surveys used a combination of 
the UV-excess and emission-line methods to discover objects.  Most of these surveys 
are substantially deeper than the Markarian survey.  More recently, the Madrid (UCM) 
survey \cite{UCM},\cite{JG} used the emission-line method, but worked in
the red to select galaxies via their H$\alpha$ emission.  A new line-selected survey 
which detects sources via both [O III]$\lambda$5007 and H$\alpha$ is described in 
detail in the following section.

\section{KISS -- KPNO International Spectroscopic Survey}

\subsection{Project Overview}

The KPNO International Spectroscopic Survey (KISS) was initiated in 1994 as
a collaborative effort between astronomers from Russia, Ukraine, and the
US who shared a common interest in the study of galaxian activity.  The
survey was envisioned as being the ``next generation Markarian survey" 
for active and star-forming galaxies.   Principal members of the KISS team
are C. Gronwall (Wesleyan), V. Lipovetsky and A. Kniazev (SAO), 
T. Boroson (NOAO), J. Moody (BYU), T. Thuan (Virginia), Y. Izotov (Ukraine),
and the author.  In addition, a number of current and former students 
have contributed a great deal to the program.  These include Wesleyan students
J. Herrero, L. Frattare, K. Kearns, K. Rhode, K. Kinemuchi, J. Lee, S. Randall, 
and N. Harrison, as well as summer interns L. Brenneman, E. Condy, and M. Santos.

KISS detects galaxies which exhibit strong emission lines.  We refer to it 
as the ``next generation" survey for extragalactic emission-line objects 
since it combines the traditional objective-prism survey method with
modern detector technology and computer-based analysis.  The survey will
discover objects with a wide range of types of galaxian activity, including 
Seyfert Galaxies (Sy 1/Sy 2), Starburst Nucleus Galaxies (SBNs), HII Galaxies, 
Blue Compact Dwarfs (BCDs), and perhaps even a few high-redshift QSOs.

The single biggest improvement that KISS makes compared to previous objective-prism 
surveys is the use of a CCD as the detector.  The advent of large format CCDs
(2k $\times$ 2k and larger), when combined with the wide-field imaging of
Schmidt telescopes, provides for the large field-of-view (FOV) capability
necessary for carrying out large areal surveys.  Previously, photographic
plates had remained the detector of choice for such surveys simply because
of their superior FOV.  However, CCDs provide a large number of advantages
over photographic plates.  These include obvious factors such as much
higher quantum efficiency, lower noise, good spectral response over the
entire optical portion of the spectrum, and large dynamic range.  In addition,
CCDs give us the ability to use automated selection methods to detect
ELGs, and allow us to quantify the selection function directly from the 
survey data.

The primary goal of KISS is to produce a high quality survey whose selection 
function and completeness limits can be accurately quantified in order to allow 
for statistical studies.  The initial hope was to be able to survey significantly 
deeper (in both flux and redshift) than previous surveys of this type.  Our 
conservative goal was to reach the equivalent of two magnitudes fainter than 
previous surveys.  All indications are that we have achieved that goal.

\subsection{Observational Technique}

The telescope used for the survey is the 24-inch Burrell Schmidt on Kitt Peak.
It is owned and operated by Case Western Reserve University; the KISS group
currently purchases time for the survey from CWRU.  The detector used for the 
first survey strip was a 2048$^2$ STIS CCD with 21 micron pixels.  The FOV of 
the CCD was 70 arcmin (1.18$^\circ$) square, and the pixel scale was 2.03 
arcsec/pixel.

Data were taken both with and without a prism.  Direct images (no prism) were 
obtained through B and V filters.  These images are used to obtain astrometry 
and photometry of all objects in each field.  Two sets of spectral (prism) data
were acquired.  The first set were taken in the blue portion of the spectrum
and utilized a 2-degree prism together with a specially designed filter covering 
4800-5500 \AA.  The second set of spectra were taken in the red with a 4-degree 
prism and a filter covering 6400-7200 \AA.  Dispersions for both prisms were 
$\sim$20 \AA /pixel.  The red spectra are used to detect objects via their H$\alpha$ 
emission, while the blue spectra detect galaxies primarily by the [O III]$\lambda$5007
line.  The use of filters to restrict the wavelength range of the spectra was
necessary to reduce the level of the night sky background in the spectral images,
and to reduce the length of the individual spectra.  Given the depth of our images,
spectral overlap in the slitless spectral images would be severe if
the length of the spectra were not restricted.

\begin{figure}[t]
\begin{center}
\leavevmode
\epsfverbosetrue
\epsfxsize=0.80\textwidth
\epsfbox{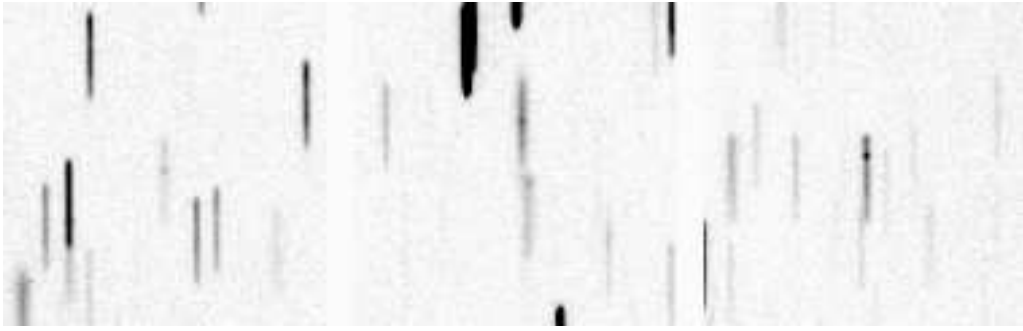}
\end{center}
\vskip -0.2in
\caption[]{Examples of 4 newly discovered ELGs detected in the KISS survey.  
{\bf Left:} An example of a fairly faint (B=18.6) ELG. {\bf Center:} Two
ELG candidates: a lower-redshift object just above center, and a high-redshift 
galaxy just below it. {\bf Right:} An intermediate-redshift object with a 
fairly strong emission line.}
\label{fig:spec1}
\end{figure}

\begin{figure}[ht]
\begin{center}
\leavevmode
\epsfverbosetrue
\epsfxsize=0.90\textwidth
\epsfbox{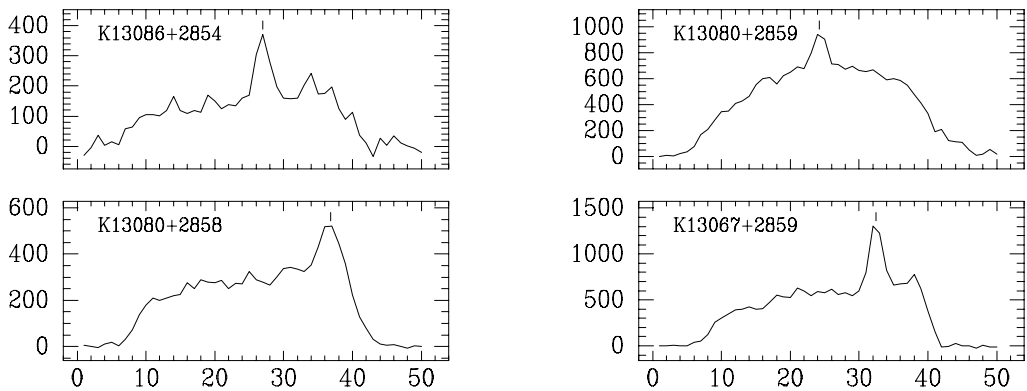}
\end{center}
\vskip -0.3in
\caption[]{Plots of the extracted spectra for the four ELGs displayed in Figure 
\ref{fig:spec1}.  K13086+2854 corresponds to the left image, K13080+2859 (upper)
and K13080+2858 (lower) to the center image, and K13067+2859 to the right.
Each spectral plot covers $\sim$800 \AA, in the range 6400 -- 7200 \AA.
Lines mark the location of the H$\alpha$ emission line.}
\label{fig:spec2}
\end{figure}

Figure \ref{fig:spec1} shows examples of four objects detected in the red spectral 
survey, while Figure \ref{fig:spec2} shows the extracted spectra of these four ELGs.  
The galaxies displayed illustrate a range of both line strength and redshift which 
is indicative of the KISS galaxies in general.

The first KISS strip was chosen to coincide with the CfA/Dartmouth Century 
Redshift survey \cite{CFA}.  This is a narrow strip one degree wide in 
declination, centered at $\delta$ = 29.5$^\circ$, and stretching 100$^\circ$ in 
right ascension, from 8$^h$ 30$^m$ to 16$^h$ 20$^m$.

\subsection{Specialized Software}

One of the important features of the KISS project is that the search process is
carried out in software.  A complete package of IRAF scripts and executables 
has been written to perform the processing required for the KISS program.  
The selection of the emission-line galaxies (ELGs) occurs in a automated 
fashion, with no tedious searches by eye required!

\begin{figure}[t]
\begin{center}
\leavevmode
\epsfverbosetrue
\epsfxsize=0.75\textwidth
\epsfbox{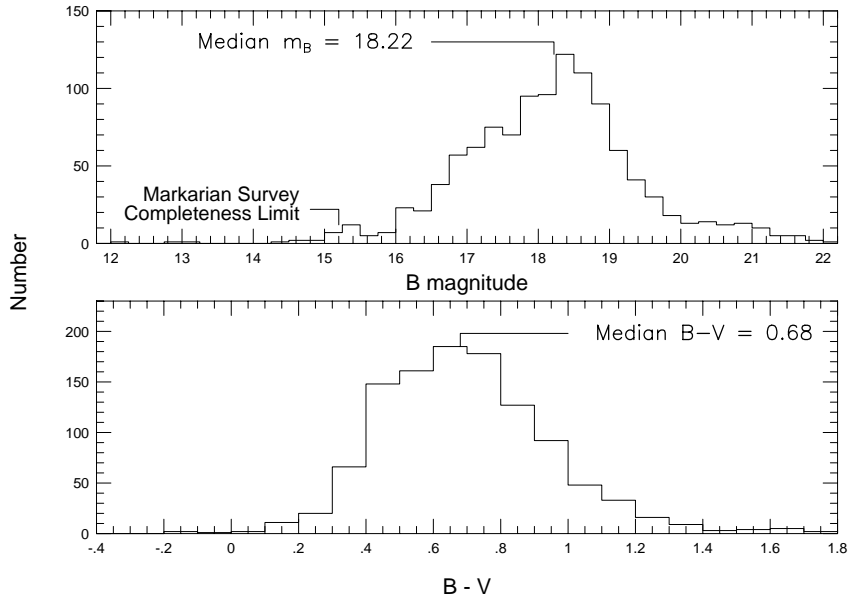}
\end{center}
\vskip -0.55in
\caption[]{Histograms showing the distribution of blue apparent magnitude
(upper) and B$-$V color (lower) for the 1126 H$\alpha$-selected KISS ELGs.
The median values for both distributions are indicated, as is the completeness 
limit of the Markarian survey (upper plot).}
\label{fig:mags}
\end{figure}

The input data for each survey field include the B and V direct images,
a deep, CR-free composite direct image created by co-adding the individual 
direct frames, the objective-prism spectral images, and data for photometric 
calibration.  The composite direct image is used in the object-detection 
phase of the analysis since it is typically 1 to 2 magnitudes deeper than 
the corresponding spectral image.  The KISS software package includes 
routines that perform the following tasks:
\smallskip
\par $\bullet$ Automatic detection of ALL objects in the direct images
\smallskip
\par $\bullet$ Digital photometry for all objects detected in the direct images
\smallskip
\par $\bullet$ Accurate astrometry for all detected objects ($\pm$0.2 arcsec)
\smallskip
\par $\bullet$ Star {\mit versus} galaxy classification for brighter sources (B $<$ 19)
\smallskip
\par $\bullet$ Coordinate transformation between the direct and spectral images
\smallskip
\par $\bullet$ Spectral extraction from 2D to 1D format for all objects
\smallskip
\par $\bullet$ Automated search for ELG candidates
\smallskip
\par $\bullet$ Measurement of redshifts and line strengths of detected ELGs
\smallskip

\par The spectral extraction step includes corrections for overlapping spectra.  
All objects detected in each field (typically 5000 -- 7000 objects) are processed
in the same way, so the processing yields photometric and astrometric information 
as well as an extracted spectrum for every source.  In a typical field the 
software selects 50 -- 100 ELG candidates, which are then checked manually.
Roughly one third of these end up in our final ELG candidate lists.  The final 
data product is an STSDAS table which contains all of the relevant information about 
each object.

\subsection{Preliminary Results}

The blue spectral portion of the first KISS strip was completed in 1996, and
consists of 102 separate fields with both direct and spectral data ($\sim$130
sq. deg.).  In the spring of 1997 we obtained red spectral data for 54 of the
102 fields, covering 68 sq. deg. in the RA range 12h 10m to 17h 0m.
Results presented here are only for the fully processed red spectral data.

The main result to report is that the survey has indeed been very successful
at finding ELGs.  In our first red spectral strip we have cataloged 1126 ELG
candidates.  This is 20.9 per field, or about {\it 16.6 per square degree}.
For comparison, the entire Markarian survey catalogued 1500 UV-excess galaxies
(0.1 galaxy per square degree), and had a completeness limit of m$_B$ = 15.2
\cite{MAZ}.  The deeper UM \cite{UM} and UCM \cite{JG} surveys both detected 
roughly 0.5 ELGs per square degree.

\begin{figure}[t]
\begin{center}
\leavevmode
\epsfverbosetrue
\epsfxsize=0.75\textwidth
\epsfbox{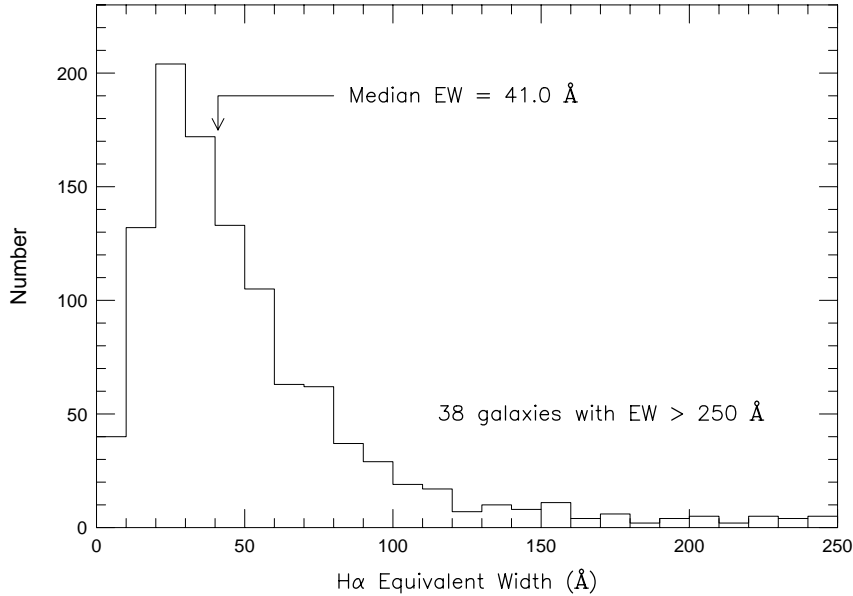}
\end{center}
\vskip -0.55in
\caption[]{Histogram showing the distribution of H$\alpha$ line equivalent
widths for the 1126 H$\alpha$-selected KISS ELGs.  The median value is indicated. 
The lines are measured directly from the objective-prism spectra, so have large
uncertainties.  The survey appears to be reasonably complete for line strengths
of 30-40 \AA\ and above.}
\label{fig:eqw}
\end{figure}

Figures \ref{fig:mags} and \ref{fig:eqw} display some of the relevant 
observational characteristics of the KISS ELG sample.  The distribution of blue 
apparent magnitude and B$-$V color is shown in Figure \ref{fig:mags}.  The 
median value of m$_B$ for the KISS galaxies
is 18.22, a full three magnitudes fainter than the completeness limit
of the Markarian survey.  The colors show a wide spread, with a median value
similar to the colors of late-type spirals.  The distribution of H$\alpha$
equivalent widths is shown in Figure \ref{fig:eqw}.  Since the galaxies are 
H$\alpha$-selected, this plot is actually more indicative of the depth of the
survey.  Although a more detailed analysis is required to quantify the 
completeness limit of the survey (e.g., Salzer \cite{S89}), it would appear that
the survey finds essentially all objects with H$\alpha$ equivalent widths 
down to 30--40 \AA.

\begin{figure}[t]
\begin{center}
\leavevmode
\epsfverbosetrue
\epsfxsize=0.75\textwidth
\epsfbox{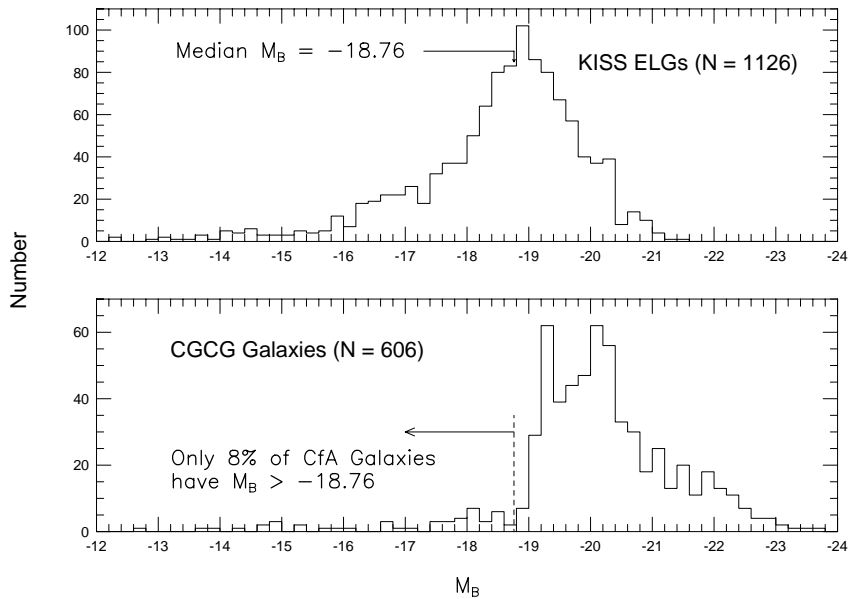}
\end{center}
\vskip -0.55in
\caption[]{Histograms showing the distribution of blue absolute magnitude for the 
1126 H$\alpha$-selected KISS ELGs (upper) and the 606 ``normal" galaxies (lower) from
the CGCG \cite{CGCG} which are located in the same area of the sky.  The median 
luminosity for the ELGs is indicated.  The KISS ELG sample is made up of predominantly
lower-luminosity galaxies, making this line-selected sample particularly powerful for 
studying dwarf galaxies.}
\label{fig:abs}
\end{figure}

\begin{figure}[ht]
\begin{center}
\leavevmode
\epsfverbosetrue
\epsfxsize=0.75\textwidth
\epsfbox{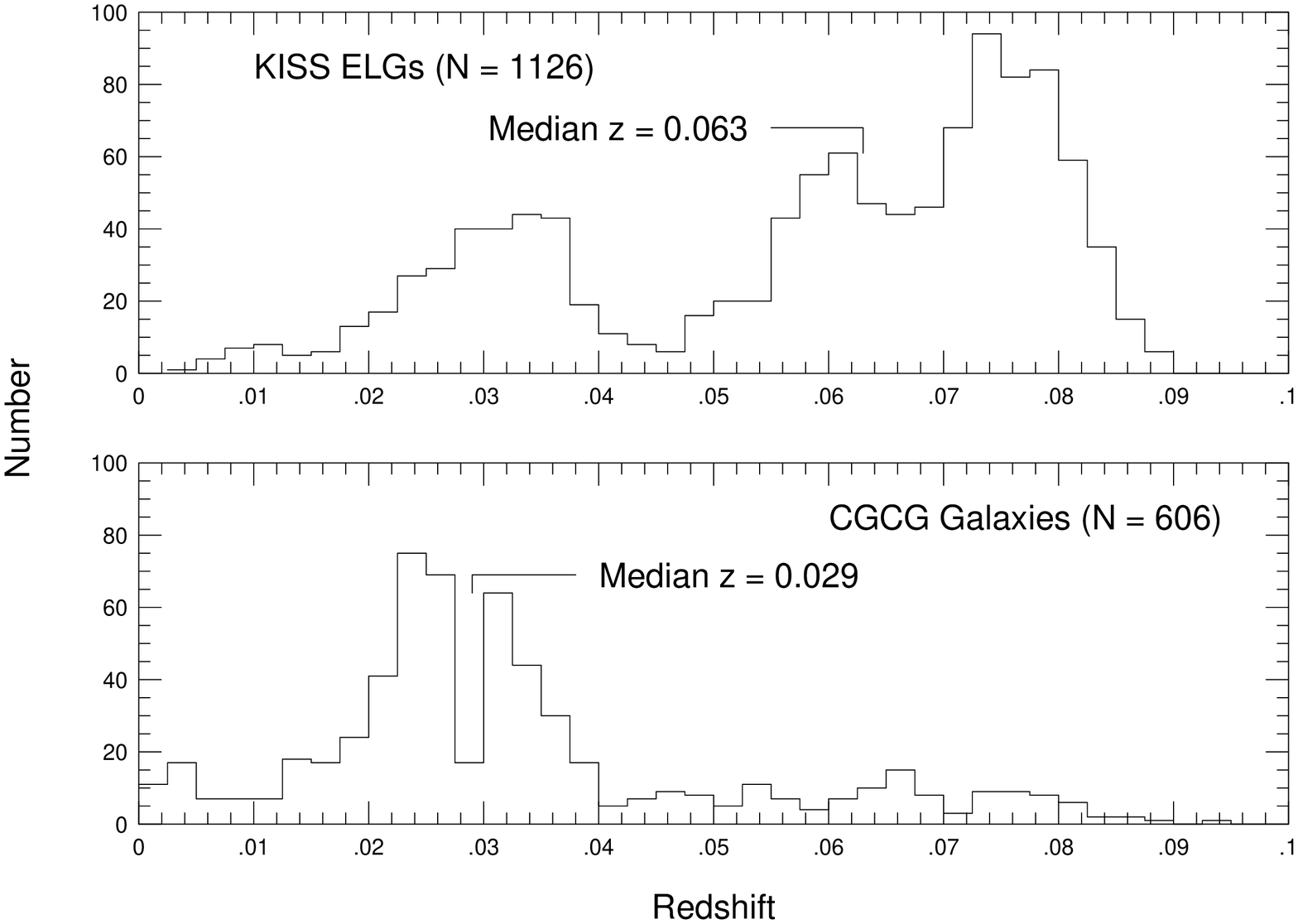}
\end{center}
\vskip -0.55in
\caption[]{Histograms showing the distribution of redshift for the 1126 
H$\alpha$-selected KISS ELGs (upper) and the 606 ``normal" galaxies (lower) from
the CGCG \cite{CGCG} which are located in the same area of the sky.  The median 
redshift is indicated in both plots.  Note that the number of KISS ELGs continues 
to rise up to the cut-off of the filter used for the survey, indicating that
the survey is volume-limited for the more luminous galaxies.  The deficit of ELGs
between z = 0.038 and 0.054 is due to a large void.}
\label{fig:zplot}
\end{figure}

It is possible to measure rough redshifts from the objective-prism spectra 
directly, and hence estimate the luminosities of the KISS ELGs.  It should be
emphasized that these redshifts are of limited precision, with an estimated accuracy 
of 700--800 km/s.  The luminosity and redshift distributions for the KISS 
galaxies are plotted in Figures \ref{fig:abs} and \ref{fig:zplot}, respectively.  
The lower panel of each figure also shows, for reference, the corresponding 
distributions for the ``normal" CGCG galaxies \cite{CGCG} 
located in the same volume of space as the ELGs.  Among the key results
illustrated here are: (1) the ELGs are dominated by intermediate- and 
low-luminosity galaxies, as is commonly found for line-selected samples of
this type; (2) the magnitude-limited CGCG sample is made up of predominantly
higher-luminosity objects; and (3) the median redshift of the KISS galaxies is
more than double that of the CGCG galaxies, and the histogram doesn't turn over
and start decreasing before the redshift limit imposed by the filter 
(z $\sim$ 0.085) is reached.  This latter point implies that the KISS ELGs 
constitute a volume-limited sample for the brighter luminosities.

More details of the survey and its constituents will be presented in
Salzer et al. \cite{S98}.

\subsection{What Can One Do With KISS?}

KISS provides a very deep, statistically complete sample of active and star-forming
galaxies.  Such a sample can be used to study a wide range of problems in modern
astrophysics.  Space doesn't allow us to go into details, but a brief listing
will be instructive.  The list can be broken up into three broad categories:

{\bf Galaxian Properties / Galaxy Evolution.}  Areas to pursue include studies 
of the frequency of AGN/starburst activity among the galaxian population, 
characteristics of the large-scale star-formation processes occurring in galaxies, 
the chemical enrichment/evolution in galaxies (particularly dwarfs), and the 
existence of a metallicity--luminosity relationship for starbursting galaxies.  
In addition, it will be fruitful to correlate the catalog of ELGs with surveys at 
other wavelengths, such as the radio, FIR, and x-ray.  The first steps of a 
correlation between the FIRST radio survey \cite{FRS} and KISS (the FIRST--KISS 
project) is already underway.

{\bf Activity in Galaxies.}   The KISS survey will provide a deep, complete sample
of AGN which will be a valuable complement to existing samples of radio and x-ray
selected Seyferts.  Among the interesting topics that could be investigated are
studies of the active stages of galaxian evolution, the physics of AGN/QSO phenomena 
with an emphasis on testing the unified model, environmental influences on activity, 
and similarities/differences between AGN selected at optical, radio, and x-ray
wavelengths.

{\bf Observational Cosmology.}   The KISS survey will contribute 
a great deal in the area of observational cosmology.   For example, pre-selecting 
galaxies for redshift survey work which exhibit optical emission lines is an 
efficient way to probe Large-Scale Structure \cite{S89}, \cite{R94}.  Studies of
the spatial distribution of dwarf galaxies \cite{S94}, \cite{P95}, \cite{T98},
made possible with dwarf-rich samples such as these, will allow for the study
of biasing in the spatial distribution of low-mass galaxies.  Our preliminary
results indicate a significant population of dwarf galaxies within nearby voids. 
The importance of this type of sample for measuring the primordial helium abundance 
has already been mentioned.  We expect to detect several dozen excellent candidates 
for future work.  Finally, the KISS sample will be useful
as a low-redshift comparison group to the galaxian population at higher redshift.
We will be able to measure more accurately than ever before the space densities of 
starbursting galaxies locally, which may well have an impact on resolving the faint 
blue galaxy problem.  KISS will also provide a more accurate estimate of the local 
star-formation rate density.  The paper by Gronwall \cite{CG} presents our preliminary 
results on this topic.

\section{Summary}

The KISS program is currently moving ahead on a number of fronts.  The 
processing of the data from our first survey strip has just been finished,
and the results are being prepared for publication.  We are also starting
additional strips in both the north and south Galactic caps, and will 
substantially increase the sky area surveyed over the next few years.
Finally, we have begun a systematic program of follow-up spectroscopy of
KISS ELG candidates.  To date we have obtained spectra for $\sim$150 ELGs
from the first red survey list.  The results are extremely encouraging, indicating
that 97\% of the KISS ELG candidates are in fact strong emission-line galaxies.
Many new Seyfert galaxies and starbursting dwarfs have already been
discovered.  The details of the follow-up spectroscopy will be reported upon 
in the literature in the near future.  

The search for and subsequent study of
low-metallicity galaxies provides many useful clues to the formation and
chemical evolution of dwarf galaxies.  In particular, the existence of a few
apparently young galaxies raises important questions regarding the
formation of galaxies in general, and may provide insights into the early
stages of the formation of the larger galaxies which dominate the cosmic
landscape today.  The KISS project is helping to address many of these questions 
by discovering large samples of dwarf galaxies.  One of our main goals is to 
discover additional examples of extreme systems such as I Zw 18, in order to 
both quantify their significance in the overall galaxian population and to 
provide new targets for study.  Perhaps of even more importance will be the 
discovery of a large, statistically-complete sample of dwarfs with a wide range 
of star-formation characteristics.  This sample will allow us to understand
more fully the role dwarf galaxies play in the star-formation rate of the local
universe, which in turn will lead to a better understanding of what 
researchers are seeing at higher redshift.


%
\vfill
\end{document}